\documentclass[12pt,floatfix,superscriptaddress,aps,prd,preprint,showpacs]{revtex4-1}
\usepackage[english]{babel}
\usepackage[utf8]{inputenc}
\usepackage[T1]{fontenc}
\usepackage{amsmath,amssymb,amsfonts,amsthm,amssymb,amsbsy,amsopn,amstext}
\usepackage[mathcal]{eucal}
\usepackage{mathrsfs}
\usepackage{latexsym}
\usepackage{bm}
\usepackage{wrapfig}
\usepackage{color}
\usepackage{units}
\usepackage{textcomp}
\usepackage{graphicx}
\usepackage{subfigure}
\usepackage{hyperref}
\usepackage{slashed}
\usepackage{float}
\graphicspath{{./fig/pdf/}}

\begin{document}

\title{Quasinormal frequencies of self-dual black holes}
\author{Victor Santos}
\email{victor\underline{\space}santos@fisica.ufc.br}
\affiliation{Universidade Federal do Cear\'a (UFC), Departamento de F\'isica, Campus do Pici, Fortaleza - CE, C.P. 6030, 60455-760 - Brazil}
\author{R. V. Maluf}
\email{r.v.maluf@fisica.ufc.br}
\affiliation{Universidade Federal do Cear\'a (UFC), Departamento de F\'isica, Campus do Pici, Fortaleza - CE, C.P. 6030, 60455-760 - Brazil}
\author{C. A. S. Almeida}
\email{carlos@fisica.ufc.br}
\affiliation{Universidade Federal do Cear\'a (UFC), Departamento de F\'isica, Campus do Pici, Fortaleza - CE, C.P. 6030, 60455-760 - Brazil}

\begin{abstract}
One simplified black hole model constructed from a semiclassical analysis of loop quantum gravity (LQG) is called self-dual black hole. This black hole solution depends on a free dimensionless parameter $P$ known as the \textit{polymeric} parameter and also on the $a_{0}$ area related to the minimum area gap of LQG.  In the limit of $P$ and $a_{0}$ going to zero, the usual Schwarzschild-solution is recovered. Here we investigate the quasinormal modes (QNMs) of massless scalar perturbations in the self-dual black hole background. We compute the QN frequencies using the sixth order WKB approximation method and compare them with numerical solutions of the Regge-Wheeler equation. Our results show that as the parameter $P$ grows, the real part of the QN frequencies suffers an initial increase and then starts to decrease while the magnitude of the imaginary one decreases for fixed area gap $a_{0}$. This particular feature means that the damping of scalar perturbations in the self-dual black hole spacetimes are slower, and their oscillations are faster or slower according to the value of $P$.
\end{abstract}

\date{\today}
\maketitle
\section{Introduction}
\label{sec-1}

General relativity (GR) is an intrinsically nonlinear theory, where to derive useful quantities from the field equations are difficult except for a few cases, for which  usually there are additional symmetries in the metric tensor. One fruitful approach is to consider the weak-field approximation, leaving the field equations more treatable for many investigations where the full equations are too difficult to solve. In particular, this approach led to the prediction of Gravitational Waves (GWs).

The emission of gravitational waves can be associated with a variety of physical processes, such as astrophysical phenomena involving the evolution of binary systems and stellar oscillations or cosmological processes that occurred in the very early universe \cite{2009-Sathyaprakash.Schutz-LRR}. Signals produced by these processes have different intensities and characteristic frequencies. In general, their spectral properties depend on the generating phenomenon \cite{2013-Riles-PPNP}.

Considerable progress has been made in the development of gravitational wave detectors. Large ground-based gravitational wave interferometers, such as LIGO \cite{1992-Abramovici.etal-S-325}, GEO-600 \cite{1997-Lueck.GEO600Team-CaQG-1471}, TAMA-300 \cite{1994-Coccia.etal-Proceedings} and VIRGO \cite{2015-FAFONE-2025} are increasing their detection rates, approaching then the original design sensitivity \cite{2014-Evans-GRG-46}. These detections will provide useful information about the composition of the astrophysical objects that generate them, like neutron stars. Moreover, one of the most interesting aspects of the detection of gravitational waves is the connection with black hole physics. The gravitational radiation emitted by an oscillating black hole carry a fingerprint that may lead to a direct observation of its existence \cite{1997-Thorne-a-arxiv}.

The investigation of the gravitational radiation from an oscillating star is usually done by studying the perturbations of the stellar or black hole spacetimes. Such radiation exhibits a certain set of characteristic frequencies, which are independent of the process that generated it. These oscillation frequencies become complex due to the emission of gravitational waves radiation, and the real and imaginary parts begin to represent the actual and damping frequencies, respectively. Such characteristic oscillations are called quasinormal modes (QNM). A striking feature of the QNM's is that they are directly connected to the parameters characterizing the black hole (mass, charge and angular momentum). The initial studies on the black hole perturbations were carried out by Regge and Wheeler concerning the stability of Schwarzschild black holes \cite{1957-Regge.Wheeler-PR-1063}, followed by Zerilli \cite{1970-Zerilli-PRD-2141,1974-Zerilli-PRD-860}. The QNM's have been found in perturbation calculations of particles falling into Schwarzschild and Kerr black holes and in the collapse of a star to form a black hole \cite{2003-Motl.Neitzke-A-307,1992-Nollert.Schmidt-PRD-2617}. Numerical investigations of the fully nonlinear equations of GR have provided results that agree with the perturbation calculations \cite{1994-Gundlach.etal-PRD-883,1999-Nollert-CaQG-159,1985-Schutz.Will-TAJ-33}.

The determination of QNM spectrum for real-world systems are very complicated
once that no analytical solutions are known for the relevant astrophysically
systems, and, therefore, the investigation must proceed numerically.
Additionally, the quasinormal frequencies form a discrete set of points scattered on the complex plane, so that an exhaustive scan of this plane is required to find the entire spectrum \cite{BlazquezSalcedo:2012pd}. Furthermore, the QNM's are known to not form a complete set of functions, in the sense that a signal can not be written for all times as a linear combination of these modes. A detailed account of quasinormal modes in asymptotically flat spacetimes, their properties, and a discussion of its incompleteness can be found in Refs. \cite{1999-Kokkotas.Schmidt-LivingRev} and \cite{1999-Nollert-CaQG-159}, and references therein.

In recent years, it has been suggested that the QNM's of black holes might play a role in quantum gravity, mainly in approaches like string theory and loop quantum gravity (LQG). Indeed, it was proposed that the study of the black hole quasinormal modes in anti-de Sitter spacetimes could be useful to determine properties of conformal field theories in the context of the AdS-CFT correspondence \cite{2000-Horowitz.Hubeny-PRD-24027}. However, such black hole solutions can not describe astrophysical effects, since they are constructed in a negative cosmological constant spacetime. In the context of loop quantum gravity, it was also suggested that the large frequency limit for the QNM's can be used to fix the Immirzi parameter, a parameter measuring the quantum of area \cite{2003-Dreyer-PRL-81301,2003-Motl-A-1135}. Nevertheless, this is a fundamental issue that remains open in this field. Inspired by these considerations, black hole quasinormal modes have been computed in various background spacetimes, and in four or higher dimensions \cite{2004-Konoplya-JoPS-93,2003-Konoplya-PRD-024018}.

The semiclassical analysis of the full LQG black hole leads to an effective self-dual metric, where the singularity is removed by replacing the singularity by an asymptotically flat region \cite{2009-Modesto.Premont-Schwarz-PRD-064041}. As this effective metric recovers Schwarzschild metric as a limit, it is compelling to investigate possible quantum modifications to already well-known effects discussed in a classical Schwarzschild metric. For example, the study of the stability properties of the Cauchy horizon in two different self-dual black hole solutions was carried out in Ref. \cite{Brown2011376}. In Ref. \cite{2012-Hossenfelder.etal-ap-1202.0412} the authors studied the wave equation of a scalar field and calculated the particle spectra of an evaporating self-dual black hole. Moreover, the thermodynamical properties of self-dual black holes, using the Hamilton-Jacobi version of the tunneling formalism were investigated in Refs. \cite{Silva2013456,Anacleto2015181}.

The present work has as its main goal to examine the quasinormal mode frequencies of a self-dual black hole, by taking a scalar field as a model of a gravitational wave. Although a realistic scenario would require the computation of the frequencies for spin-2 perturbations, scalar perturbations are a good starting point. They are simpler to study, and the gravitational field itself contains scalar degrees of freedom \cite{1957-Regge.Wheeler-PR-1063}, providing a qualitative similar analysis. We computed the frequencies employing a 3rd-order and 6th-order WKB approximation, comparing them with previous calculations in the classical limit. We also solved the full Regge-Wheeler equation in the time domain, comparing the semianalytic frequencies as a fit for the time profile. We found that the polymeric parameter $P$ induces a more slowly decay for the scalar field while its oscillation becomes higher or lower according to the value of $P$. 

The paper is structured as follows. We introduce self-dual black hole solution in section \ref{sec-2}. In section \ref{sec-3} we compute the self-dual black hole quasinormal frequencies, exhibiting the time-domain profile and finally we summarize the work and present our final remarks in section \ref{sec-4}.

\section{Self-dual black hole}
\label{sec-2}

Loop Quantum Gravity \cite{2004-Ashtekar.Lewandowski-CaQG-53,2007-HAN.etal-IJoMPD-1397,1993-Thiemann.Kastrup-N-211} provided a description of the universe in its early stages. One key feature is the resolution of initial singularity \cite{2009-Ashtekar-G-707,2004-Bojowald.Morales-Tecotl-LP-421}, where the Big Bang singularity is replaced by a quantum bounce. A black hole in this framework was constructed from a modification of the holonomic version of the Hamiltonian constraint \cite{2009-Modesto.Premont-Schwarz-PRD-064041}. This kind of black hole, called self-dual black hole, has a solution which depends on a parameter $\delta$, called \emph{polymeric parameter}, which labels elements in a class of Hamiltonian constraints. These constraints are compatible with spherical symmetry and homogeneity, and they can be uniquely fixed from asymptotic boudary conditions, yielding the proper classical Hamiltonian in the limit $\delta\to 0$.

Following the approach of Refs. \cite{Brown2011376,2012-Hossenfelder.etal-ap-1202.0412}, the loop quantum corrected Schwarzschild black hole can be described by the effective metric
\begin{equation}
\label{eq:self-dual-metric}
{\textrm{d}s}^{2} = -F(r){\textrm{d}t}^2 + \frac{\textrm{d}r^2}{G(r)}
+ H(r){\textrm{d}\Omega_2}^2,
\end{equation}
where ${\textrm{d}\Omega_2}^2 = {\textrm{d}\theta}^2 + \sin^2{\theta}{\textrm{d}\phi}^2$ and the metric functions are given by
\begin{align}
F(r) &= \frac{(r-r_{+})(r-r_{-})(r+r_{\ast})^{2}}{r^{4}+a_{0}^{2}},\\
G(r) &= \frac{(r-r_{+})(r-r_{-})r^{4}}{(r+r_{\ast})^{2}(r^{4}+a_{0}^{2})},\\
H(r) &= \frac{a_{0}^{2}}{r^{2}} + r^{2},
\end{align}
with $r_+ = 2m$, $r_{-}=2mP^2$, $r_{\ast}=2mP$. The parameter $m$ is related to the ADM mass $M$ by $M=m(1 + P)^{2}$. Moreover, the function $P$ is called \emph{polymeric function}, and it is related to the polymeric parameter $\delta$ by $P(\delta)={(\sqrt{1+\gamma^2\delta^{2}}-1)}/{(\sqrt{1+\gamma^2\delta^2}+1)}$, where $\gamma$ denotes the Barbero-Immirzi parameter. The area gap $a_0$ equals to $A_{\textrm{min}}/8\pi$, with $A_{\textrm{min}}$ being the minimum area of LQG, namely
\begin{equation}
A_{\textrm{min}} = 8\pi\ell_{\textrm{P}}^2\gamma\sqrt{j_{\textrm{min}}(j_{\textrm{min}}+1)},
\end{equation}
where $\ell_{\textrm{P}}$ is the Planck length and $j_{\textrm{min}}$ is the smallest value of the representation on the edge of the spin network crossing a surface. A common choice is to consider representations of the $SU(2)$ group, which leads to $j_{\textrm{min}}=1/2$ and then $a_0=\gamma\sqrt{3}\ell_{\textrm{P}}^2/2$. In this work we will assume $\gamma\sim 1$, so that we fix $a_0=\sqrt{3}/2$ in Planck units.

The self-duality property can be expressed by saying that the metric \eqref{eq:self-dual-metric} is invariant under the transformations
\begin{equation}
r\to a_0/r,\quad t\to tr_{\ast}^2/a_0,\quad r_{\pm}\to a_0/r_{\mp}.
\end{equation}
The minimal surface element is obtained when the dual coordinate $\tilde{r}=a_0/r$ takes the value $r_{\textrm{dual}}=\tilde{r}=\sqrt{a_0}$. The property of self-duality removes the black hole singularity by replacing it with another asymptotically flat region. For the dynamical aspects of this solution we refer the reader to Ref. \cite{2012-Hossenfelder.etal-ap-1202.0412}. However, the polymerization of the Hamiltonian constraint in the homogeneous region is inside the event horizon. Therefore, the physical meaning of the solution when the metric is analytically continued to all spacetime persist as an open problem \cite{Brown2011376}. Moreover, self-dual black holes have two horizons -- an event horizon and a Cauchy horizon. Cauchy horizons are unstable, and then it is not clear whether the solution has a stable interior. Nevertheless, this solution can be useful as a first approximation to the behavior of a system in a quantum gravitational framework. One advantage of this approach is that although the full black hole solution can only be presented in a numerical form, this solution has a closed form which makes our investigation easier.

Concerning the experimental bound of the polymerization parameter, a recent study based on observation of solar gravitational deflection of radio waves leads to the constraint \cite{2015-Sahu.etal-PRD-63001}
\begin{equation}
\delta\lesssim 0.1
\end{equation}
for the polymerization parameter, which implies $P\sim 10^{-3}$. We stress that this estimate has as asumption $\gamma<0.25$, and since we did a different choice for the value of the Barbero-Immirzi parameter we will therefore impose a less restrictive range of values $P<1$.

\section{Quasinormal frequencies}
\label{sec-3}

In this section we are interested in obtaining the solutions of wave equations in the presence of a background spacetime described by \eqref{eq:self-dual-metric}, and thus to find the QNM frequencies. In most cases it is not feasible to solve the wave equations and find frequencies exactly. Among the analytical methods used in literature, the WKB method is one of the most popular. The WKB method was first used by Iyer and Will \cite{iyer_black-hole_1987,1987-Iyer-PRD-3632}, and later improved to sixth order by Konoplya \cite{2004-Konoplya-JoPS-93}.

Now, we consider the behavior of a scalar field in a self-dual black hole background. The propagation of a massless scalar field is described by the Klein-Gordon equation in curved space-time
\begin{equation}
\label{eq:kg-equation}
\frac{1}{\sqrt{-g}}\partial_{\mu}\big(g^{\mu\nu}\sqrt{-g}\partial_{\nu}\Phi\big) = 0,
\end{equation}
where the field is treated as an external perturbation. In other words, we ignore the backreaction. Spherical symmetry allow us to decompose the field in terms of spherical harmonics,
\begin{equation}
\label{eq:spherical-harmonics-decomposition}
\Phi(t,r,\theta,\varphi) =\sum_{\ell=0}^{\infty}\sum_{m=-\ell}^{\ell} \frac{1}{\sqrt{H(r)}}\Psi_{\ell,m}(t,r) Y_{\ell,m}(\theta,\varphi),
\end{equation}
and substituting Eq. \eqref{eq:spherical-harmonics-decomposition} into Eq. \eqref{eq:kg-equation} one can find the radial coefficients which satisfy the equation
\begin{equation}
\label{eq:regge-wheeler-equation}
-\frac{\partial^{2}\Psi}{\partial t^{2}}
+\frac{\partial^{2}\Psi}{\partial x^2} + V(x)\Psi = 0,
\end{equation}
where we introduced the tortoise coordinate $x$, defined by
\begin{equation}
\frac{\textrm{d}x}{\textrm{d}r} = \frac{r^4 + a_0^2}{r^2(r-r_-)(r-r_+)}.
\end{equation}
After integrating, the tortoise coordinate assumes the explicit form
\begin{align}
x&= r-\frac{a_0^2}{r r_- r_+}-\frac{\left(a_0^2+r_-^4\right) \log \left|
  r-r_-\right|}{r_-^2 \left(r_+-r_-\right)}\nonumber\\
& +\frac{\left(a_0^2+r_+^4\right) \log \left|r-r_+\right|}{r_+^2 \left(r_+-r_-\right)}+\frac{a_0^2\left( r_- + r_+\right) \log (r)}{r_-^2 r_+^2},
\end{align}
such that $x\rightarrow\pm\infty$ for $r\rightarrow r_{\mp}$. The function $V(x)$ is called \emph{Regge-Wheeler potential}, and it encodes the information regarding the black hole geometry. To the present case it is given by \cite{Brown2011376}
\begin{align}
\label{potV}
V(r(x)) &= \frac{(r-r_-) (r-r_+)}{(r^4 + a_{0}^2)^4}\Big[ r^2 \Big(a_{0}^4 \left(r
   \left( \, \left(K^2-2\right) r+r_- + r_+\right)+2 K^2 r r_{\ast}+K^2 r_{\ast}^2 \right) \nonumber \\
   & +2 a_{0}^2 r^4 \Big(\left(K^2+5\right) r^2+2 K^2 r r_{\ast}+K^2 r_{\ast}^2-5 r (r_-+r_+) +5 r_-   r_+ \Big)\nonumber\\
   & +r^8\left(K^2 (r+r_{\ast})^2+r (r_-+r_+)-2 r_- r_+\right)\Big)  \Big],
\end{align}
with $K^2=\ell(\ell+1)$. From this expression, we note that $V(x)$ vanishes at $r=r_{-}$ and $r=r_{+}$, and as discussed in Ref. \cite{Brown2011376}, the maximum of the potential occurs at $r\approx \ell_{\textrm{P}}/3$.

We will now look for stationary solutions in the form $\Psi\sim \psi(r)e^{-\text{i}\omega t}$. With this ansatz, equation \eqref{eq:regge-wheeler-equation} takes the form of a Schrödinger equation given by
\begin{equation}
\label{eq:schrodinger-equation}
\frac{\partial^{2}\psi}{\partial x^2} - \big[\omega^2 - V(x)\big]\psi = 0.
\end{equation}
Before solving Eq. \eqref{eq:schrodinger-equation}, we must state the boundary conditions. Since we are dealing with a black hole geometry, physically acceptable solutions are purely ingoing near the horizon in the form
\begin{equation}
\label{eq:ingoing-solutions}
\psi^{\textrm{in}}(r)\sim
\begin{cases}
e^{-\text{i}\omega x}&(x\to -\infty)\\
C_\ell^{(-)}(\omega)e^{-\text{i}\omega x}+C_\ell^{(+)}(\omega)e^{\text{i}\omega x}&(x\to +\infty),
\end{cases}
\end{equation}
where $C_{\ell}^{(-)}(\omega)$ and $C_{\ell}^{(+)}(\omega)$ are complex constants that depend on the variables $\ell$ and $\omega$.

QNMs of black holes are in general defined as the set of frequencies $\{\omega_{\ell n}\}$ such that $C_{\ell}^{(-)}(\omega_{\ell n})=0$, i.e., which are associated with purely outgoing wave at the spatial infinity and purely ingoing wave at the event horizon. Here the $\ell$ and $n$ labels are integers, called multipole and overtone numbers respectively. Since the spectrum of QNMs is in fact determined by the eigenvalues of the equation \eqref{eq:schrodinger-equation}, it is convenient to promote the analogy with quantum mechanics and make use of semianalytical methods. Next, we will employ a high-order WKB approach to calculating the QNM frequencies for some values of $\ell$ and $n$ indexes.
\subsection{WKB analysis}
\label{sec-3-1}
The WKB approximation method to evaluate the QNMs was first implemented by Schutz and Will in the calculation of the particles scattering around black holes \cite{1985-Schutz.Will-TAJ-33}. An improvement in the method up to sixth WKB order has been given by Konoplya \cite{2004-Konoplya-JoPS-93,2003-Konoplya-PRD-024018}. The method can be applied if the potential has the form of a barrier, going to constant values when $x\to\pm\infty$. At each limit, the frequencies are calculated by matching the power series of the solution near the maximum of the potential through its turning points.

In the general case,  Konoplya's WKB formula to the QN frequencies can be written as
\begin{equation}
\label{eq:quasinormal-frequencies}
\frac{\text{i}(\omega_n^2 - V_0)}{\sqrt{-2V''_0}} - \sum_{i=2}^{6}\Lambda_{i} =
n+\frac{1}{2},
\end{equation}
where $V^{\prime\prime}_{0}$ is the second derivative of the potential on the maximum $r_{0}$ and $\Lambda_i$ are constant coefficients which depend on the effective potential and its derivatives (up to 12-th order) at the maximum.

\begin{table}[htb]
\caption{\label{tbl:qnms-wkb3-lqg}Quasinormal modes computed using 3rd-order WKB approximation, for different values of the polymerization parameter. The multipole number is fixed to $\ell=1$.}
\centering
\begin{tabular}{rccc}
\hline
$P$ & $\omega_{0}$ & $\omega_{1}$ & $\omega_{2}$\\
\hline
0.1 & $0.3031-0.0981\textrm{i}$ & $0.2704-0.3079\textrm{i}$ & $0.2249-0.5293\textrm{i}$\\
0.2 & $0.3174-0.0971\textrm{i}$ & $0.2887-0.3030\textrm{i}$ & $0.2487-0.5195\textrm{i}$\\
0.3 & $0.3291-0.0944\textrm{i}$ & $0.3047-0.2928\textrm{i}$ & $0.2706-0.5006\textrm{i}$\\
0.4 & $0.3376-0.0900\textrm{i}$ & $0.3176-0.2775\textrm{i}$ & $0.2891-0.4730\textrm{i}$\\
0.5 & $0.3421-0.0839\textrm{i}$ & $0.3260-0.2573\textrm{i}$ & $0.3028-0.4374\textrm{i}$\\
0.6 & $0.3417-0.0763\textrm{i}$ & $0.3288-0.2330\textrm{i}$ & $0.3095-0.3950\textrm{i}$\\
0.7 & $0.3355-0.0677\textrm{i}$ & $0.3246-0.2058\textrm{i}$ & $0.3071-0.3483\textrm{i}$\\
0.8 & $0.3227-0.0587\textrm{i}$ & $0.3121-0.1783\textrm{i}$ & $0.2940-0.3016\textrm{i}$\\
\hline
\end{tabular}
\end{table}

\begin{table}[htb]
\caption{\label{tbl:qnms-wkb6-lqg}Quasinormal modes computed using 6th-order WKB approximation, for different values of the polymerization parameter. The multipole number is fixed to $\ell=1$.}
\centering
\begin{tabular}{rccc}
\hline
$P$ & $\omega_{0}$ & $\omega_{1}$ & $\omega_{2}$\\
\hline
0.1 & $0.3056-0.0983\textrm{i}$ & $0.2770-0.3073\textrm{i}$ & $0.2399-0.5401\textrm{i}$\\
0.2 & $0.3194-0.0973\textrm{i}$ & $0.2940-0.3025\textrm{i}$ & $0.2606-0.5283\textrm{i}$\\
0.3 & $0.3307-0.0946\textrm{i}$ & $0.3088-0.2924\textrm{i}$ & $0.2796-0.5073\textrm{i}$\\
0.4 & $0.3388-0.0901\textrm{i}$ & $0.3204-0.2771\textrm{i}$ & $0.2951-0.4784\textrm{i}$\\
0.5 & $0.3430-0.0839\textrm{i}$ & $0.3283-0.2568\textrm{i}$ & $0.3073-0.4400\textrm{i}$\\
0.6 & $0.3424-0.0763\textrm{i}$ & $0.3309-0.2323\textrm{i}$ & $0.3139-0.3944\textrm{i}$\\
0.7 & $0.3358-0.0677\textrm{i}$ & $0.3244-0.2063\textrm{i}$ & $0.3026-0.3558\textrm{i}$\\
0.8 & $0.3233-0.0587\textrm{i}$ & $0.3156-0.1764\textrm{i}$ & $0.3061-0.2903\textrm{i}$\\
\hline
\end{tabular}
\end{table}

In tables \ref{tbl:qnms-wkb3-lqg} and \ref{tbl:qnms-wkb6-lqg} we present the
tabulated quasinormal frequencies using respectively the 3-rd and 6-th order WKB
method for a fixed multipole $\ell=1$. Some comments are in
  order. First, the QN frequencies associated with the scalar field have a
  negative imaginary part; this means that the QNMs decay exponentially in time,
  losing energy in the form of scalar waves. This result was already expected
  and has also been found for scalar, electromagnetic and gravitational
  perturbations in the Schwarzschild-like geometry \cite{Konoplya:2011qq}. They also have a similar behavior to the usual case ($P\to 0$ and
  $a_{0}\to 0$)  regarding the overtone number scale, so that the real part decreases, whereas the magnitude of imaginary one grows for crescent values of overtone number $n$. Our main result concerns the behavior of QN frequencies with respect to the polymeric parameter $P$. We note that the real part of the QN frequencies begins to grow and then decreases as $P$ increases, while the magnitude of the imaginary part decreases. This means that as the polymerization parameter gets larger the frequency of the massless scalar wave shift higher or lower, and it has a slower damping.

\begin{figure}[!ht]
\centering
\includegraphics[width=0.7\textwidth]{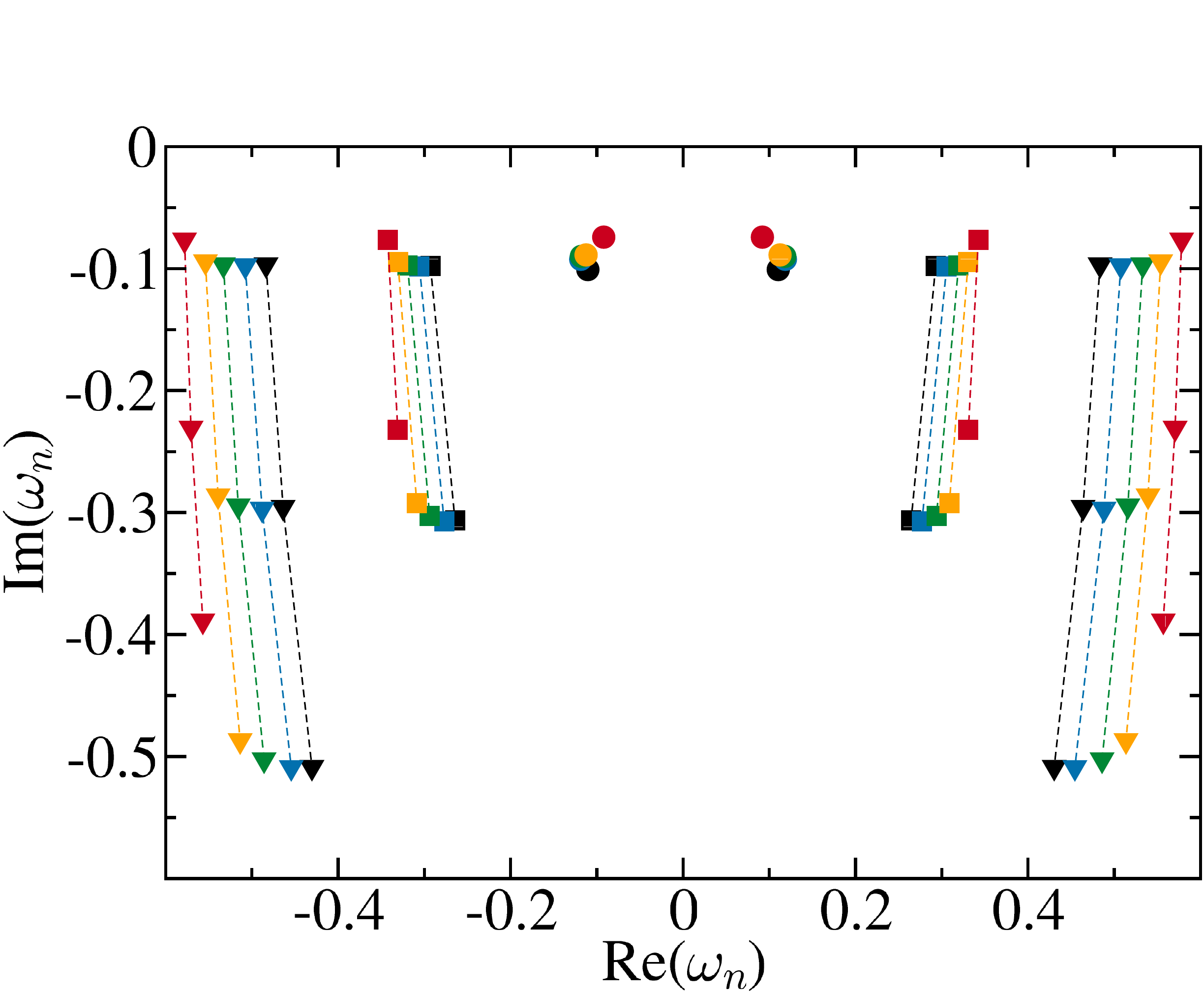}
\caption{\label{fig:wkb6-qnm-freqs}Scalar field normal modes. The markers denote the multipole number as: $\ell=0$ (circle), $\ell=1$ (square), $\ell=2$ (triangle), while the colors denote the value of the polymerization parameter: $P=0.1$ (blue), $P=0.2$ (green), $P=0.3$ (orange), $P=0.6$ (red). The black markers denote the frequencies for the Schwarzschild black hole.}
\end{figure}

It is interesting to plot the frequencies in the complex plane in contrast with the frequencies of the Schwarzschild case. In figure \ref{fig:wkb6-qnm-freqs}  this is done considering three families of multipoles $\ell=0,1,2$ for the Schwarzschild case. We can notice the deviation from the classical case as we increase the polymerization function. Looking at the right side of the figure, we conclude that the frequency curves are moving counterclockwise as $P$ grows, and this twisting effect becomes more apparent for larger values of the angular quantum number $\ell$. Furthermore, these results differ from the frequencies found in Ref. \cite{saleh_quasinormal_2014}, where a quantum correction was introduced as a deformation as introduced in Ref. \cite{KAZAKOV1994153}.

\begin{figure}[!ht]
\centering
\includegraphics[width=0.8\textwidth]{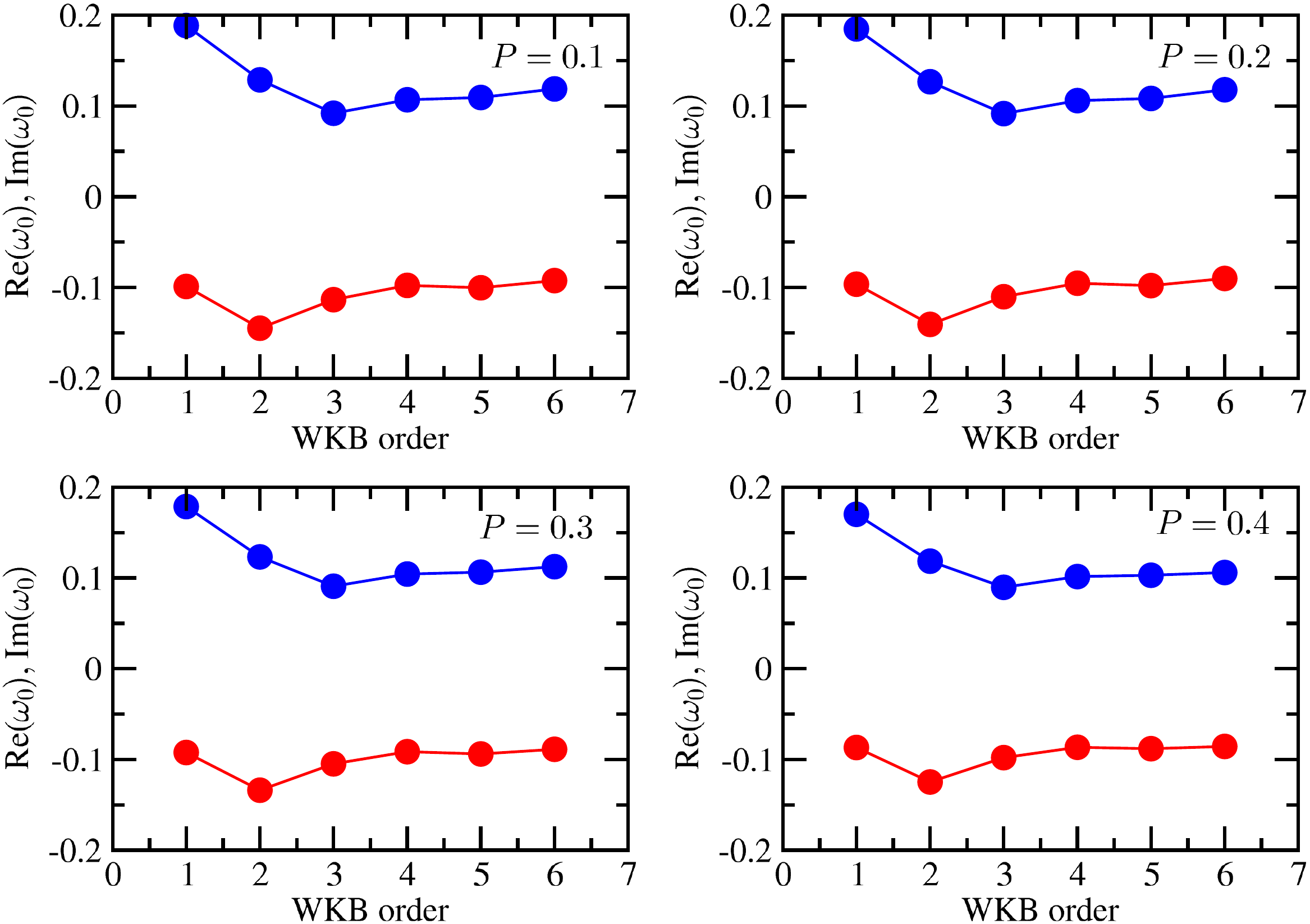}
\caption{\label{fig:wkborder}Real (blue/top) and imaginary (red/bottom) parts of the quasinormal frequencies as a function of the WKB order for the $\ell=0$, $n=0$ mode, for some values of the polymerization parameter $P$.}
\end{figure}

The 6th-order WKB formula usually provides a better approximation than the 3rd-order formula. Since the WKB series converges only asymptotically, in general we can not guarantee the convergence only by increasing the WKB order. In Fig. \ref{fig:wkborder} we present the real and imaginary parts of the frequencies as a function of the WKB order for fixed multipole and overtone, respectively $\ell=0$ and $n=0$. From this we can observe that indeed both real and imaginary parts converge as we increase the order, evidencing the effectiveness of the method.

The presented results one show that the real part of the quasinormal frequencies  first increase and then start to decrease with the polymerization parameter, meaning a shift from blue to red into the spectrum of the massless scalar wave function. The magnitude of the imaginary part decreases with $P$, meaning a slower damping than of the classical result. The conclusion is that loop quantum black holes decay slower than its classical analogue. In this context, we expect that this is caused by the domination of curvature effects in the quantum regime.

\subsection{Time-domain solution}
\label{sec-3-2}

Due to the complex form of the effective potential, in order to contemplate the role of the QN spectrum in a time-dependent scattering we will investigate the scalar perturbations in the time domain. For this we employ the characteristic integration method developed by Gundlach and collaborators \cite{1994-Gundlach.etal-PRD-883}. This method uses light-cone coordinates $u=t-x$ and $v=t+x$, in a way the wave equation can be recasted as
\begin{equation}
\label{eq:rw-eq-characteristic}
\bigg(4\frac{\partial^2}{\partial u\partial v} + V(u, v)\bigg)\Psi(u,v) = 0.
\end{equation}

Equation \eqref{eq:rw-eq-characteristic} can be integrated numerically by a simple finite-difference method, using the discretization scheme
\begin{equation}
\Psi(N) = -\Psi(S) + \Psi(W) + \Psi(E) - \frac{h^2}{8}V(S)\big[\Psi(W) + \Psi(E)\big] + \mathcal{O}(h^{4}),
\end{equation}
where $S=(u,v)$, $W=(u+h,v)$, $E=(u,v+h)$, $N=(u+h,v+h)$ and $h$ is an overall grid scale factor.

The initial data are specified at null surfaces $u=u_0$ and $v=v_0$. We chose a
Gaussian profile centered at $v=v_{c}$ and width $\sigma$ on $u=u_0$,
\begin{equation}
\Psi(u=u_0, v) = Ae^{-(v - v_{\ast})^{2}/(2\sigma^2)},\quad \Psi(u, v_0) = \Psi_0,
\end{equation}
and at $v=v_0$ we posed a constant initial condition: $\Psi(u,v_0)=\Psi_0$. We can choose without loss of generality $\Psi_0=0$. Once chosen the null data, integration proceeds at $u=\textrm{const.}$ lines in the direction of increasing $v$.

\begin{figure}[!ht]
\center
\subfigure[ref1][\label{fig:solution-l-0-log} $\ell=0$.]{\includegraphics[width=7.6cm]{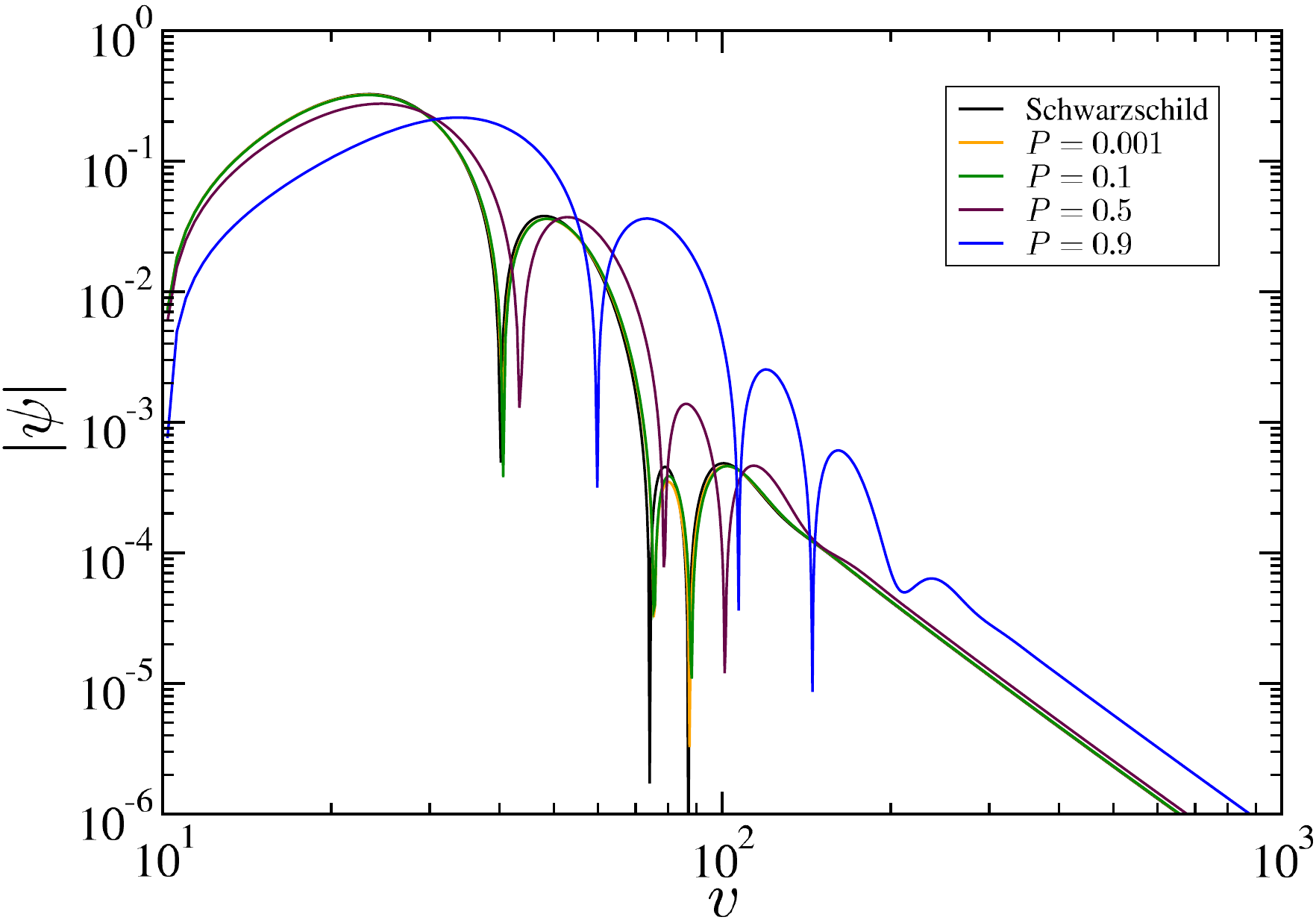}}
\qquad
\subfigure[ref2][\label{fig:solution-l-1-log} $\ell=1$.]{\includegraphics[width=7.6cm]{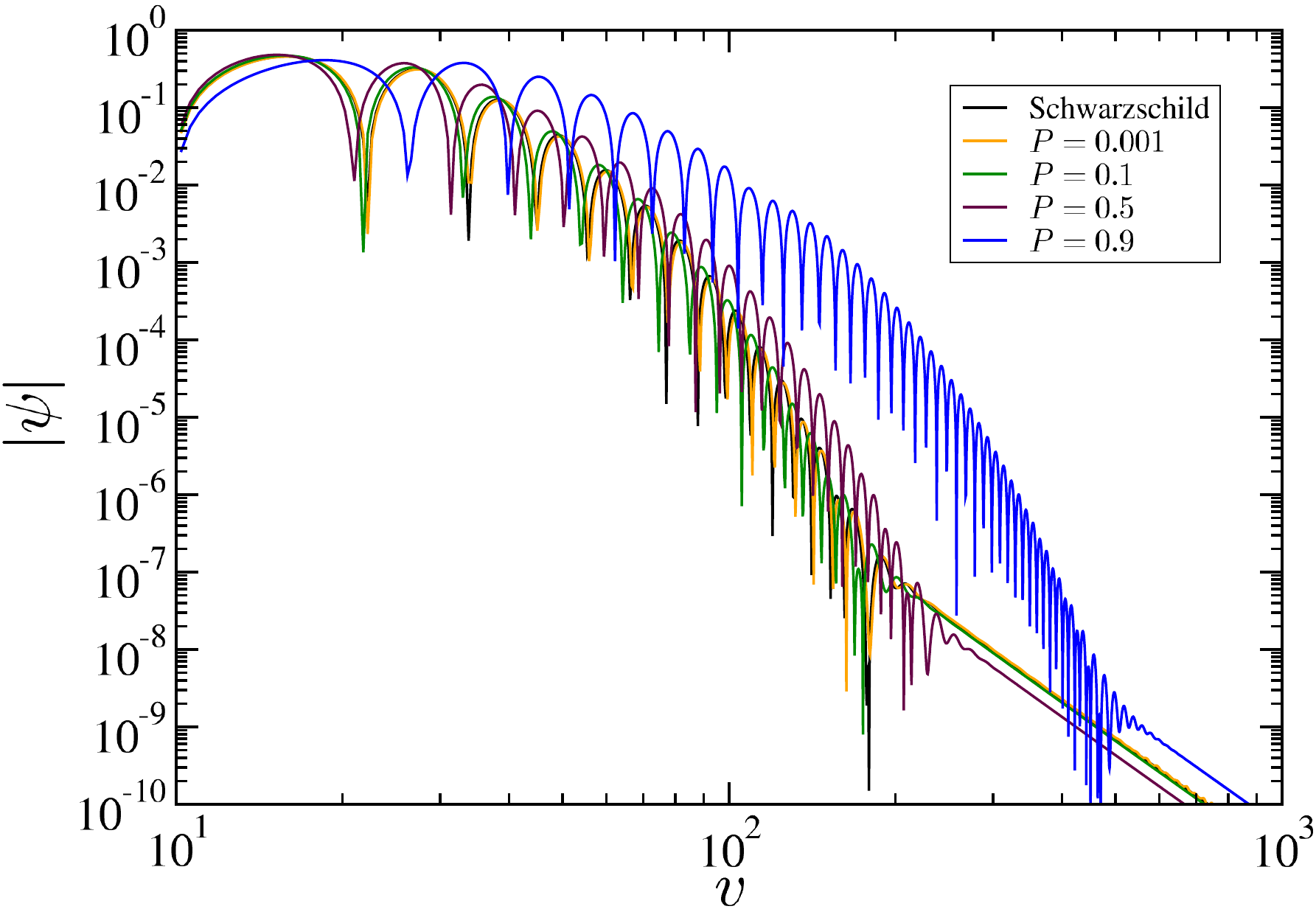}}\\
\qquad
\subfigure[ref3][\label{fig:solution-l-2-log} $\ell=2$.]{\includegraphics[width=7.6cm]{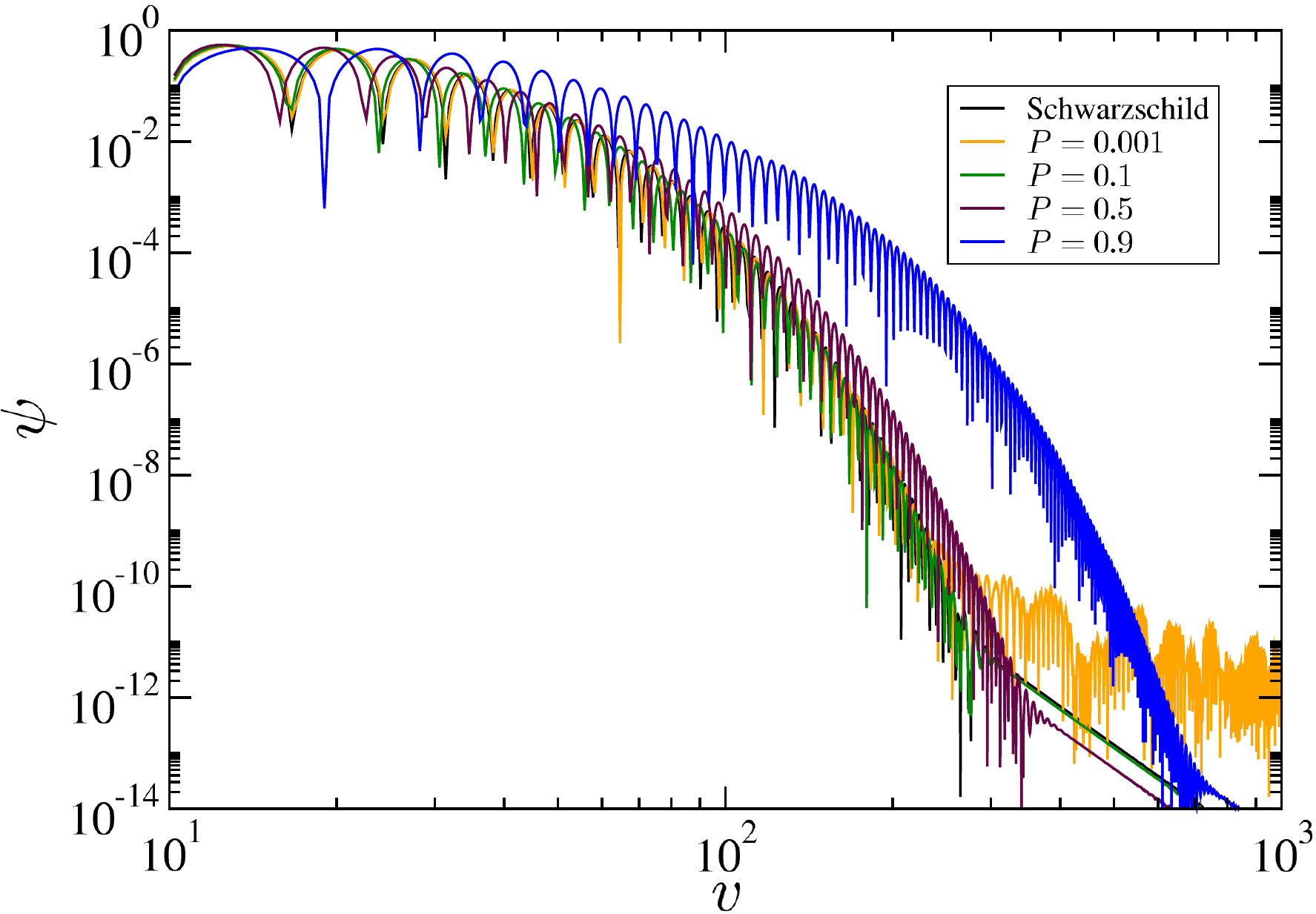}}
\caption{\label{fig:solution-log}Log-log plots of the time domain profiles for the scalar perturbations at $x=10M$, for several values of the polymerization parameter $P$.}
\end{figure}

We now report our results for the scalar test field. We set $m=1$ without loss of generality, and we chose as null data a Gaussian profile with width $\sigma=3.0$ centered at $u=10$ at the surface $u=0$, and $\Psi_0=0$. The grid was chosen to be $u\in[0,1000]$ and $v\in[0,1000]$, with points sampled in such a way we have the overall grid factor $h=0.1$.

Figures \ref{fig:solution-l-0-log}, \ref{fig:solution-l-1-log}, and \ref{fig:solution-l-2-log} show the typical evolution profiles, in comparison with that of the classical Schwarzschild black hole. The classical result is visually indistinguishable from the profile $P=0.001$, which is consistent with the fact that $P\to 0$ corresponds to a classical profile. We can observe the domination of damped oscillations (quasinormal ringing) in the region $v\sim 200$, before a power-law tail takes over in later times. As we consider larger multipole numbers, we observe a discrepancy between the quantum and classical results. indeed, with $\ell=2$ the curve $P=0.001$ (Fig. \ref{fig:solution-l-2-log}) seems to have a larger oscillation even during the tail, while the other curves started to be dominated by the power-law tail. This is still consistent with the conclusion obtained from the analysis of the WKB spectrum presented in Sec. \ref{sec-3-1}, since althought we took a small value for $P$ we should still consider the fluctuations due to the area gap $a_0$ in the scattering potential.

\begin{figure}[!ht]
\centering
\includegraphics[width=0.75\textwidth]{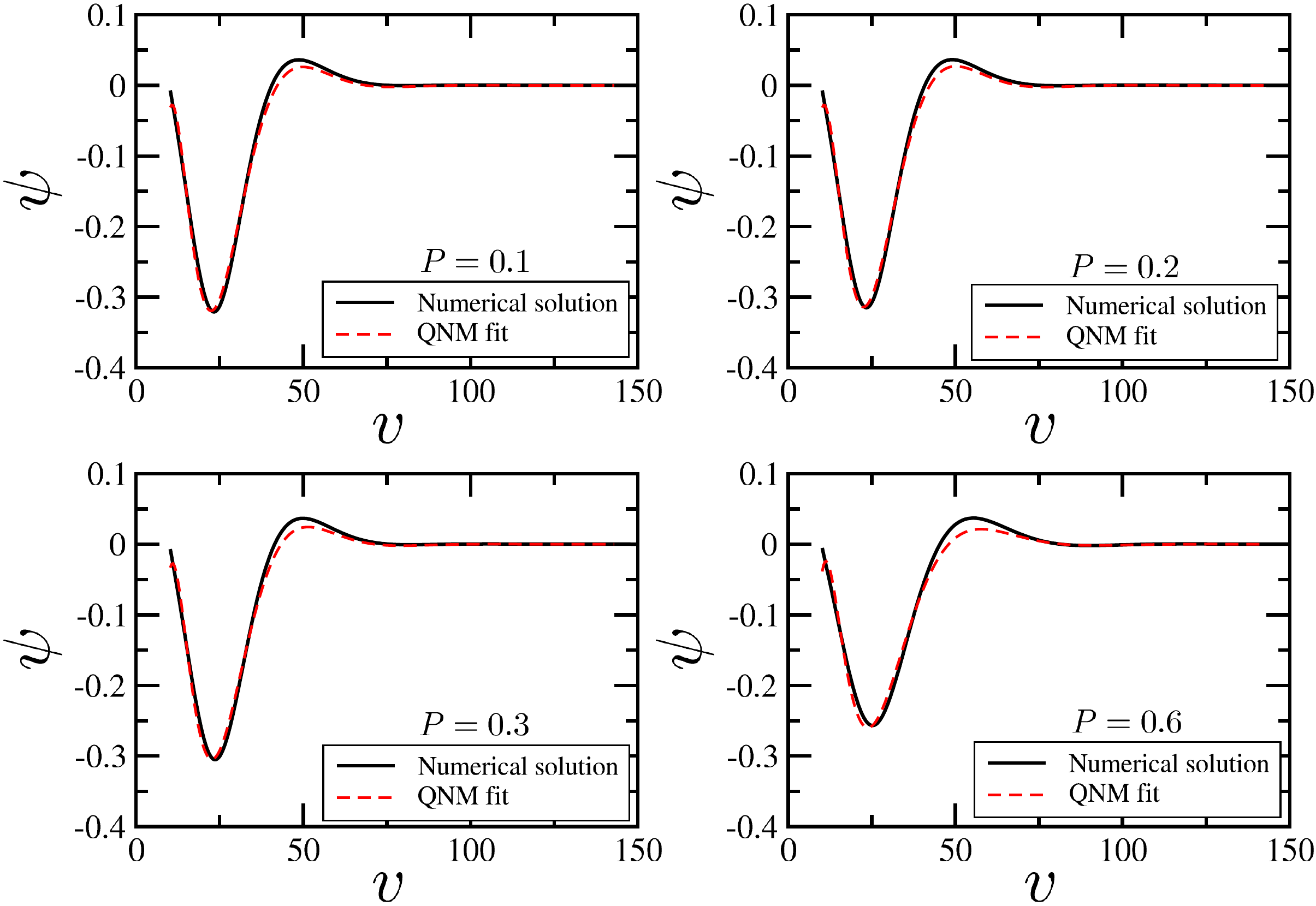}
\caption{\label{fig:wkb6-best-fit-l-0}Numerical evolution of waveform (solid line) for $\ell=0$ and the fit from the 6-th-order WKB frequencies, for several values of the polymerization parameter. For the fit we took the fundamental mode and the first overtone.}
\end{figure}

\begin{figure}[!ht]
\centering
\includegraphics[width=0.75\textwidth]{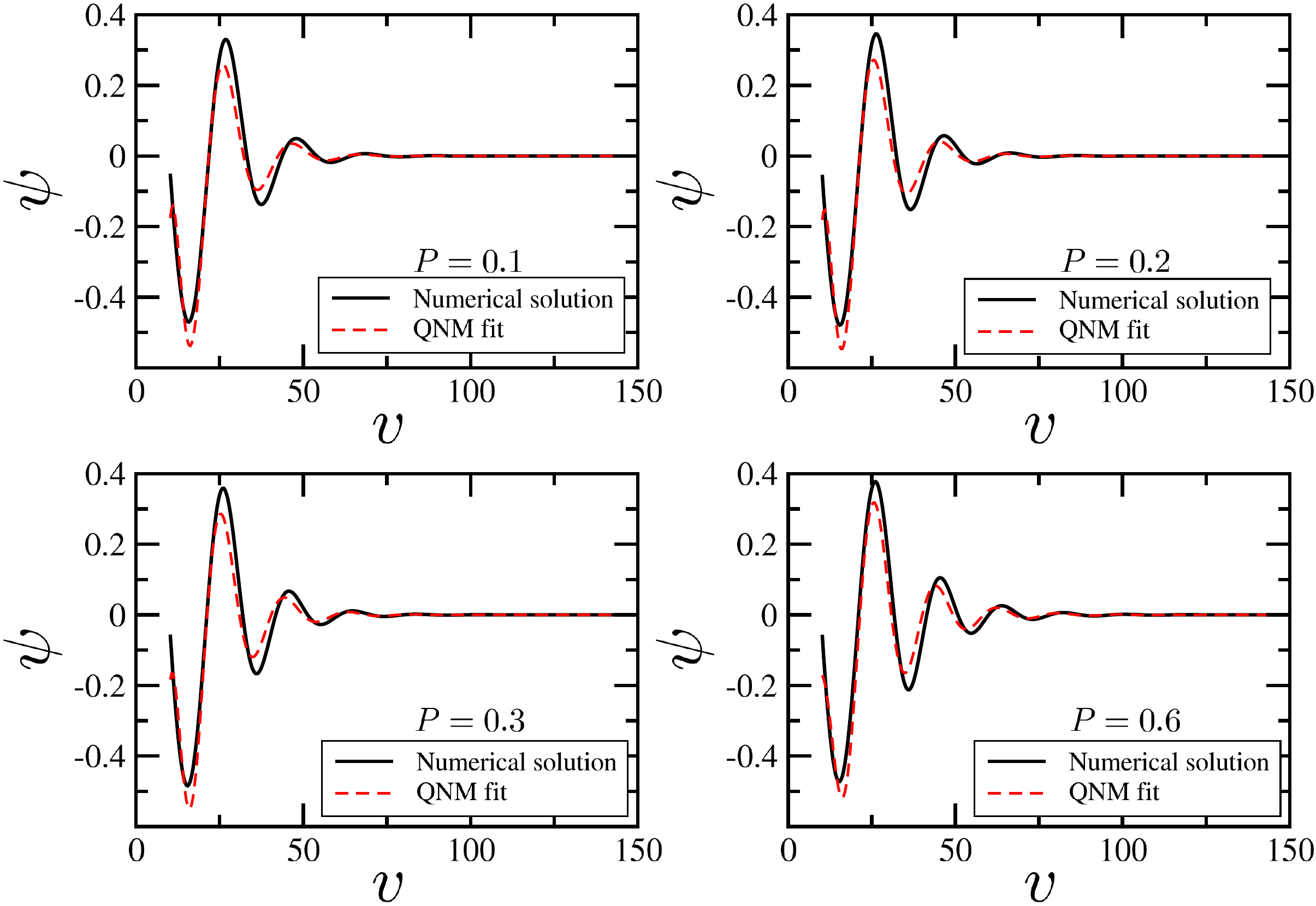}
\caption{\label{fig:wkb6-best-fit-l-1}Numerical evolution of waveform (solid line) for $\ell=1$ and the fit from the 6-th-order WKB frequencies, for several values of the polymerization parameter. For the fit we took the fundamental mode and the first overtone.}
\end{figure}


In Figs. \ref{fig:wkb6-best-fit-l-0} and \ref{fig:wkb6-best-fit-l-1}, we compared the time-domain solutions with a solution constructed from a linear combination of plane waves whose frequencies are given by the WKB method estimates. We considered a linear combination of the fundamental mode and the first overtone only. We can observe that the time-domain profiles agree with the WKB solution with good accuracy, even when considering only a small set of overtones. From the time-profile, the frequencies can be obtained using the Prony method. We did not so, since the Prony method requires knowledge from the interval of quasinormal ringing, for which there is no general method to obtain, and also due the fact that it is hard to obtain modes beyond the fundamental one. As a matter of fact, those modes could be easier determined by the WKB approximation.

\section{Conclusions}
\label{sec-4}

In this work, we presented an investigation on the quasinormal frequencies of a scalar field perturbations around of the self-dual black hole. This black hole solution is interesting because some properties induced by LQG effects can be inferred from this model \cite{Brown2011376}. We evaluated the QN frequencies using both 3rd-order and 6th-order WKB approximation, and the convergence of the method was established by considering the lower order approximations. We solved the Regge-Wheeler equation corresponding to the scalar perturbations around the black hole, verifying the agreement with the QNM frequencies from the WKB method.

The computed frequencies implied a slower decaying rate for black holes in quantum regime. This behaviour has been shown to be sensitive to the polymerization parameter and to the area gap, converging to the classical behavior when these parameters vanishes. We hope this could give clues about the evaporation process of quantum black holes at qualitative level.

We emphasize that our analysis has the limitation that the metric used does not correspond to the full LQG solution. Also, we considered only scalar perturbations. In a more realistic scenario one should treat gravitational (tensorial) perturbations as well. However QNMs from scalar perturbations of variations of non-charged, non-spinning black holes are widely studied in the literature and the QNMs found here are able to be compared with those arising in these other models. Furthermore, the qualitative relationship we have found is expected to carry over to realistic, astrophysical black holes.

\begin{acknowledgments}
This work was partially supported by the Brazilian agencies Coordenação de Aperfeiçoamento de Pessoal de Nível Superior (CAPES) and Conselho Nacional de Desenvolvimento Científico e Tecnológico (CNPq). Victor Santos would like to thank R. Konoplya for providing the WKB correction coefficients in electronic form. He also would like to thank Prof. Abhay Ashtekar for the kind hospitality at  Institute of Gravitation and Cosmos of The Pennsylvania State University, where part of this work was done.
\end{acknowledgments}


%

\end{document}